 \documentclass[prl,amsmath,amssymb,twocolumn, showpacs, superscriptaddress,10pt]{revtex4-1}

\usepackage{amsmath}
\usepackage{hyperref}
\usepackage{graphicx}
\usepackage{amsfonts}
\usepackage{amsthm}
\usepackage{cases}
\usepackage{bm}

\usepackage{color}
\definecolor{Blue}{rgb}{0.00, 0.00, 1.00}
\definecolor{Red}{rgb}{1.00, 0.00, 0.00}

\newcommand{\blue}{\color{Blue}}

\hypersetup{
    colorlinks=true,       
    linkcolor=red,          
    citecolor=blue,        
    filecolor=magenta,      
    urlcolor=cyan           
}

\newcommand{\nn}{\nonumber}
\newcommand{\be}{\begin{equation}}
\newcommand{\ee}{\end{equation}}

\newcommand{\beqn}{\begin{eqnarray}}
\newcommand{\eeqn}{\end{eqnarray}}

\DeclareMathOperator{\Li}{Li}

\DeclareMathOperator{\Tr}{Tr}
\DeclareMathOperator{\erfc}{erfc}
\DeclareMathOperator{\erf}{erf}

\newcommand{\moy}[1]{\ensuremath{\langle #1 \rangle}}

\newcommand{\var}[1]{{\rm Var}\left(#1\right)}

\begin{document}

\title{Entanglement Entropy and Full Counting Statistics for $2d$-Rotating Trapped Fermions}

\author{Bertrand Lacroix-A-Chez-Toine}
\affiliation{LPTMS, CNRS, Univ. Paris-Sud, Universit\'e Paris-Saclay, 91405 Orsay, France}
\author{Satya N. \surname{Majumdar}}
\affiliation{LPTMS, CNRS, Univ. Paris-Sud, Universit\'e Paris-Saclay, 91405 Orsay, France}
\author{Gr\'egory \surname{Schehr}}
\affiliation{LPTMS, CNRS, Univ. Paris-Sud, Universit\'e Paris-Saclay, 91405 Orsay, France}

\date{\today}

\begin{abstract}
We consider $N$ non-interacting fermions in a $2d$ harmonic potential of trapping frequency $\omega$ and in a rotating frame at angular frequency $\Omega$, with $0<\omega - \Omega\ll \omega$. At zero temperature, the fermions are in the non-degenerate lowest Landau level and their positions  
are in one to one correspondence with the eigenvalues of an $N\times N$ complex Ginibre matrix. For large $N$, the fermion density is uniform over the 
disk of radius $\sqrt{N}$ centered at the origin and vanishes outside this disk. We compute exactly, for any finite $N$, the R\'enyi entanglement entropy of order $q$, $S_q(N,r)$, as well as the cumulants of order $p$, $\moy{N_r^{p}}_c$, of the number of fermions $N_r$ in a disk of radius $r$ centered at the origin. For $N \gg 1$, in the (extended) bulk, i.e., for $0 < r/\sqrt{N} < 1$, we show that $S_q(N,r)$ is proportional to the number variance $\var{N_r}$, despite the non-Gaussian fluctuations of $N_r$. This relation breaks down at the edge of the fermion density, for $r \approx \sqrt{N}$, where we show analytically that $S_q(N,r)$ and $\var{N_r}$ have a different $r$-dependence.   
\end{abstract}


\maketitle

The last decade has seen tremendous progress in the experimental manipulation of cold atoms~\cite{BDZ08}. For Fermi gases, the recent development of Fermi quantum microscopes \cite{Fermicro1,Fermicro2,Fermicro3} allows to take a true instantaneous ``picture'' of the gas. Thanks to these techniques, the {\it Full Counting Statistics} (FCS), i.e., the statistics of the number $N_{\cal D}$ of particles in a given domain ${\cal D}$, is now accessible experimentally. In particular, an important physical observable is the number variance ${\rm Var}(N_{\cal D}) = \langle N_{\cal D}^2\rangle - \langle N_{\cal D}\rangle^2$. The FCS is thus a fundamental tool to characterise the (quantum and thermal)
fluctuations in many-body systems and it has thus been studied in various contexts, ranging from quantum transport in mesoscopic physics \cite{Lev96} to electronic correlations in quantum dots \cite{Been06,Gus06} or noise measurements in interacting Bose gases~\cite{Nat08}.

Another important quantity in many-body quantum systems is the bipartite entanglement entropy of ${\cal D}$ with its complement $\overline{\cal D}$. During the last ten years, entanglement entropy has generated a lot of interest, in particular because it can be used to identify critical and topological phases of matter~\cite{SRev}. 
Quantum entanglement is commonly quantified through the R\'enyi entanglement entropy, parameterised by $q \geq 1$, defined as 
\be
S_q(N,{\cal D})=\frac{1}{1-q}\ln\Tr[\rho_{\cal D}^q]\;{\rm with}\;\rho_{\cal D}=\Tr_{\overline{\cal D}}[\rho]\;,
\ee
where $\rho$ is the full density matrix of the system and $\rho_{\cal D}$, obtained by tracing out over the complement $\overline{\cal D}$, is the reduced density matrix. 
For $q\to 1$, it gives the von Neumann entanglement entropy $S_{\rm vN}(N,{\cal D})=\lim_{q\to 1}S_q(N,{\cal D}) = - \Tr \rho_{\cal D} \ln \rho_{\cal D}$. Powerful methods, including one-dimensional conformal field theory \cite{CC04,Dubail} as well as renormalisation group techniques \cite{RM04, MFS09, DMH17}, have been developed to compute $S_q(N,{\cal D})$ in various systems.

These two observables, ${\rm Var}(N_{\cal D})$ and $S_q(N,{\cal D})$, have a priori no reasons to be related to each other. However, interestingly, 
in a translationally invariant free Fermi gas in $d$ dimensions, they were found to be proportional to each other~\cite{Kli06,KL09,Hur11,V12,CMV11,CMV12_1,CLM15}. This raised the interesting possibility of extracting experimentally the entanglement entropy (which is very difficult to access experimentally) via measuring the number variance. Since this proportionality is rather crucial, it is natural to ask whether it continues to hold for a non-interacting Fermi gas, even in the presence of a confining trap that breaks the translational invariance. Indeed the trap creates an edge in space beyond which the particle density vanishes. Characterising the effect of confinement on $S_q(N, {\cal D})$, as well as on the FCS, in particular for a domain ${\cal D}$ close to the edge where fluctuations are typically strong, is a difficult and challenging issue \cite{us_PRL, us_EPL,us_PRA}. Recent analytical progresses were achieved for non-interacting Fermi gases trapped in a one-dimensional harmonic potential \cite{V12,CLM15}, exploiting a connection with the Gaussian Unitary Ensemble of random matrix theory (RMT) \cite{Ricardo,Eis13,Ricardo_PRE}. In the presence of a trap, the density has a finite support on $[-r_{\rm e}, + r_{\rm e}]$ with two edges at $\pm r_{\rm e} =\sqrt{2N}$ (in dimensionless units). The entanglement entropy $S_q(N,{\cal D})$ of the interval ${\cal D}=[-\ell,\ell]$ 
was analytically studied in the large $N$ limit. Near the trap center, for $\ell = {\cal O}(1/\sqrt{N})$, i.e., of the order of the inter-particle distance, the particles do not feel the potential curvature and on this length scale, they behave like free fermions. In this bulk regime, for $1/\sqrt{N} \ll \ell \ll \sqrt{N} $, it was shown \cite{CMV11,CMV12_1,CLM15} that, at leading order for large $N$, $S_q(N,{\cal D})$ is proportional to the number variance $\var{N_{\cal D}} = \langle N_{\cal D}^2\rangle - \langle N_{\cal D}\rangle^2 \simeq (1/\pi^2) \ln(\ell \sqrt{N})$, 
\be\label{S_var}
S_q(N,{\cal D})\approx \frac{\pi^2}{6}\left(1+\frac{1}{q}\right) \var{N_{\cal D}}\;.
\ee
In Ref. \cite{CLM15}, it was demonstrated that this relation (\ref{S_var}) extends far beyond the bulk regime, up to $\ell = {\cal O}(\sqrt{N})$, for $0<\ell/\sqrt{N}<\sqrt{2}$, even though the particles feel the confining potential on this length scale. In fact, one can show that this relation (\ref{S_var}) between entanglement entropy and the number variance holds more generally provided the fluctuations of $N_{\cal D}$ are Gaussian \cite{Hur11,CMV12_1,CLM15}. It is well known that the fluctuations of $N_{\cal D}$ are indeed Gaussian in the bulk of $1d$ harmonically trapped fermions \cite{CL95,Ricardo}, but this ceases to be true at the edge, for $\ell - r_{\rm e} = {\cal O}(N^{-1/6})$. In this regime, the number variance \cite{Ricardo} and the entropy \cite{CLM15} are given by rather complicated formal expressions and, unfortunately, it is very difficult to conclude whether or not there exists a relation between the entanglement entropy and the number variance, similar to (\ref{S_var}), at the edge~\cite{CLM15}.
\begin{figure}[h]
\includegraphics[width=\linewidth]{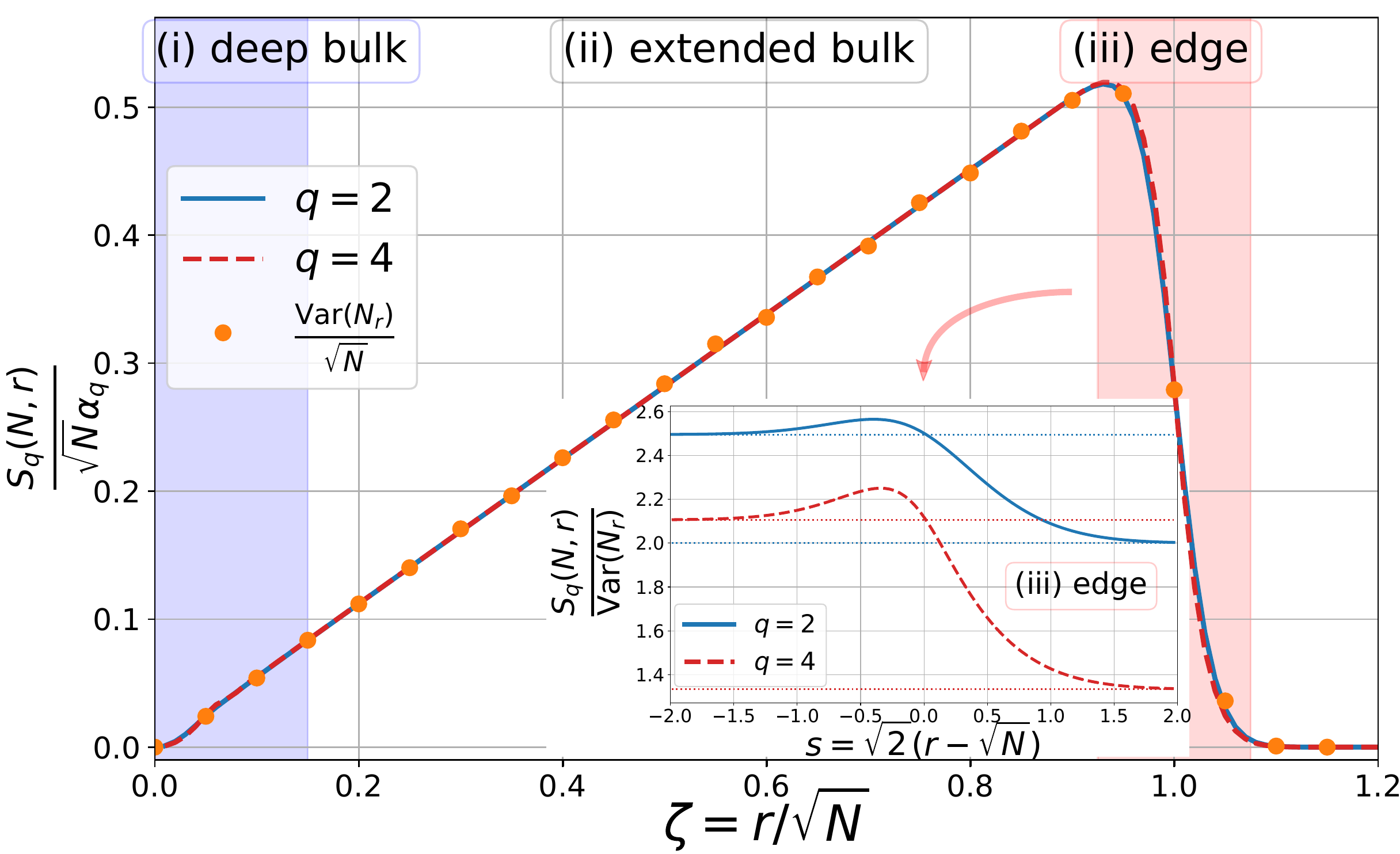}
\caption{Plot of the rescaled R\'enyi entanglement entropy $S_q(N,r)/(\sqrt{N}\alpha_q)$, with $\alpha_q$ given in (\ref{prop}) for $q=2$ (blue) and $q=4$ (dashed red) and rescaled variance $\var{N_r}/\sqrt{N}$ (orange dots) obtained numerically by diagonalisation of $5\times 10^4$ complex Ginibre matrices of linear size $N=200$ as a function of the radius $\zeta=r/\sqrt{N}$ in the extended bulk ($0<\zeta<1$). In this regime, all functions converge to the same scaling function $\zeta/\sqrt{\pi}$.
{\bf Inset}: Plot of the ratio $S_q(N,r)/\var{N_r}$ for $q=2$ (blue) and $q=4$ (dashed red) as a function of the rescaled position $s=\sqrt{2}(r-\sqrt{N})$. For $s\to -\infty$, it goes to the constant (extended bulk) value $\alpha_q$ (\ref{prop}) while for $s\to +\infty$ it goes to $q/(q-1)$.}\label{Fig_S_var}
\end{figure}

In this Letter we study non-interacting fermions in the two-dimensional $xy$ plane in the presence of a harmonic potential $V({\bf r}) = m\omega^2 r^2/2$, with $r = |{\bf r}|$, and at zero temperature. But here, at variance with previous studies, we consider the situation where the whole system is in rotation at angular frequency $\Omega<\omega$ around the $z$-axis, perpendicular to the $xy$ plane. Such rotating quantum systems have received a lot of attention, both theoretically \cite{HC00,Ho01,ABD05,TWC06} and experimentally~\cite{SCEMC04, ZASSK05} for cold atoms~\cite{Coo08}. In particular, for higher $\Omega\lesssim \omega$, they exhibit a strong analogy with quantum Hall systems \cite{Coo10}. Very little is known about the entanglement and FCS in such systems. Here, using a connection with RMT, namely the so-called Ginibre ensemble \cite{Ginibre,Mehta,Forrester_book}, we obtain exact analytical results for $S_q(N,{\cal D})$ as well as for the full statistics of $N_{\cal D}$ for a spherical domain ${\cal D}$ and show that they display a very rich behaviour. This constitutes one rare example where these observables can be studied analytically at the edge of the Fermi gas. 

{\it Model.} In the rotating frame at angular frequency $\Omega$, the many-body Hamiltonian $\hat{\cal H}_N$ is time-independent and given by $\hat{\cal H}_N=\sum_{i=1}^N \hat{H}_i$ where the single-particle Hamiltonian $\hat H_i = \hat H(\hat {\bf p}_i, \hat {\bf r}_i)$ reads
\be\label{Ham}
\hat{H}(\hat {\bf p}, \hat {\bf r}) = \frac{{\bf \hat p}^2}{2m} + \frac{m \omega^2}{2} r^2 - \Omega \hat L_z \;, 
\ee
where $\hat L_z = i \hbar(y\partial_x - x \partial_y)$ is the $z$-component of the angular momentum. The last term in (\ref{Ham}) comes from the fact that we describe the system in the rotating frame (see e.g. \cite{Landau,Leggett}). In the following, we work with dimensionless quantities and set $\omega, \hbar \omega$ and $\sqrt{\hbar/(m\omega)}$ as units of frequency, energy and length respectively. The Hamiltonian can be diagonalised explicitly by the following set of eigenfunctions $\psi_{n_1, n_2}$ indexed by two quantum numbers $n_1, n_2 = 0, 1, 2, \ldots$ (see e.g. \cite{HC00, ABD05}), 
\be\label{eigenfunction}
\psi_{n_1, n_2}(x,y) = A\, e^{\frac{r^2}{2}}(\partial_x + i \partial_y)^{n_1} (\partial_x - i \partial_y)^{n_2} e^{-r^2}
\ee
with corresponding eigenvalues 
\begin{eqnarray}\label{eigenvalues}
E_{n_1, n_2} = 1 + (1-\Omega)n_1 + (1+\Omega)n_2 \;,
\end{eqnarray}
where $A \equiv A_{n_1, n_2}$ in (\ref{eigenfunction}) is a normalizing constant, and $\Omega$ is evaluated in units of $\omega$, with $0 \leq \Omega \leq 1$. Note that this basis of eigenfunctions (\ref{eigenfunction}) is different from the standard one built from one-dimensional Hermite polynomials. Of course, in the limit $\Omega = 0$, Eq. (\ref{eigenvalues}) yields back the spectrum of the standard $2d$-harmonic oscillator. In the other limit, 
$\Omega = 1$, the energy levels $E_{n_1, n_2}$ in (\ref{eigenvalues}) depend only on $n_2$ and are thus infinitely degenerate. 
They correspond to the well known Landau levels, studied extensively in the context of quantum Hall systems. However, when $\Omega < 1$ this degeneracy, within each $n_2$-sector, is lifted. 

Here we consider the ground state of $N$ non-interacting fermions with such single-particle energy levels (\ref{eigenvalues}) and assume that $\Omega$ is sufficiently close to $1$ such that the $N$ lowest energy levels correspond to $n_2=0$ (the lowest Landau level) and $n_1=0,1, \cdots, N-1$. This indeed corresponds to the many-body ground state, as long as $E_{N-1,0} < E_{0,1}$, i.e. $1>\Omega > 1-2/N$ (note that, for bosons, a value $\Omega \approx 0.99$ was experimentally reached in \cite{SCEMC04}). In the sector $n_2 = 0$, we relabel, for convenience, the eigenfunctions $\psi_{n_1,0}$ in Eq.~(\ref{eigenfunction}) by $k = n_1 + 1 = 1, 2, \cdots, N$, and write $\psi_{n_1,0}(x,y) = \phi_{k=n_1+1}(z)$ 
\be\label{wf}
\phi_k(z)=\frac{z^{k-1}}{\sqrt{\pi \Gamma(k)}}e^{-\frac{1}{2}|z|^2}\;,\;z=x+iy\;,
\ee
with corresponding eigenvalues $E_{k-1,0}$ (\ref{eigenvalues}). The many-body ground-state wave function $\Psi_0(z_1, \cdots, z_N)$ is given by the Slater determinant constructed from the $N$ first single particle wave-functions (\ref{wf}), i.e., $\Psi_0(z_1, \cdots, z_N) = \det_{1\leq j,k \leq N}\phi_k(z_j)/\sqrt{N!}$. It is easy to see from (\ref{wf}) that $\Psi_0$ can be written in terms of a Vandermonde determinant. Consequently, the quantum joint probability distribution function (PDF) of the positions is given by
\begin{equation}\label{P_joint}
|\Psi_0(z_1,\cdots,z_N)|^2
=\frac{1}{Z_N}\prod_{i< j}|z_i-z_j|^2 e^{-\sum_{k=1}^N |z_k|^2}\;,
\end{equation}
where $Z_N$ is a normalization constant. Eq. \eqref{P_joint} coincides with the joint PDF of the eigenvalues of random matrices belonging to the 
complex Ginibre ensemble \cite{Ginibre,Mehta,Forrester_book}. This is a well known ensemble in RMT consisting of $N \times N$ matrices whose entries, both real and imaginary parts, are independent Gaussian random variables, of zero mean and variance $1/2$~\cite{Ginibre}. It is well known in the literature that the Ginibre ensemble is related to the system of $N$ fermions in a $2d$-plane and in the presence of a magnetic field perpendicular to that plane (see e.g. \cite{GNV02}). In this case, the energy levels are given by Eq.~(\ref{eigenvalues}) with $\Omega = 1$ and therefore the lowest Landau level, corresponding to $n_2 = 0$, is infinitely degenerate, i.e. all the $\phi_k = \psi_{k-1,0}$'s have the same energy. Consequently, all the Slater determinants built from $\phi_{n_1}, \phi_{n_2}, \cdots, \phi_{n_N}$ for any $n_1<n_2<\cdots<n_N$ correspond to quantum states with the same energy. Hence, there is no physical reason, even at $T=0$, to consider only the $N$-particle state built from $\phi_1, \phi_2, \cdots, \phi_N$ which is actually the only one that is related to 
the Ginibre ensemble. Instead, in the case of rotating fermions at $T=0$ considered here (\ref{Ham}), the degeneracy in the lowest Landau level (i.e., $n_2 = 0$) is lifted for $\Omega < 1$ [see Eq. (\ref{eigenvalues})] and the state built from $\phi_1, \cdots, \phi_N$ is the unique ground-state of the system. Therefore the connection to the Ginibre in Eq.~(\ref{P_joint}) comes out very naturally.

Exploiting this connection with the complex Ginibre ensemble, we immediately find that the density of fermions $\rho_N(z)$ converges in the large $N$ limit to the celebrated Girko's law \cite{Girko} $\rho_N(z)\to \rho_{\rm b}\Theta(\sqrt{N}-|z|)$, where $\Theta(x)$ is the Heaviside step-function, with uniform density $\rho_{\rm b}=1/\pi$. Hence, the typical interparticle distance is ${\cal O}(1)$. For large $N$, one thus naturally distinguishes three different regions depending on $r=|z|$ (see Fig. \ref{Fig_S_var}): (i) the {\it deep bulk} for $r = {\cal O}(1)$, (ii) the {\it extended bulk} for $0<r/\sqrt{N}<1$ and (iii) the {\it edge} for $|r-\sqrt{N}| = {\cal O}(1)$, i.e. the scale over which the density vanishes~\cite{Forrester_book}. 

{\it Main results}. 
In this Letter, we compute exactly, for any $N$, the R\'enyi entanglement entropy $S_q(N,r)$ \eqref{S_N} of the disk ${\cal D}_r=\lbrace|z|\leq r\rbrace$ as well as all the cumulants $\moy{N_r^p}_c$ (\ref{cumul_N}) of any order $p$, of the number of fermions $N_r \equiv N_{{\cal D}_r}$ in ${\cal D}_r$. 
For $N \gg 1$, we show that $S_q(N,r)$ takes different scaling forms in the three regions (i)-(iii) (see Fig. \ref{Fig_S_var})    
\be\label{S}
S_q(N,r)\approx\begin{cases}
S_q^{\rm b}\left(r\right)\;,&r = {\cal O}(1)\\
\sqrt{2} \sigma_q\,r \;,& 0<r/\sqrt{N} < 1\\
\sqrt{2N} S_q^{\rm e}\left(\sqrt{2}(r-\sqrt{N})\right),&|\sqrt{N}-r|={\cal O}(1),
\end{cases}
\ee
where the scaling functions $S_q^{\rm b,e}$ and the amplitude $\sigma_q$ can be computed explicitly \cite{Supp_mat}. In particular  the scaling function at the edge reads (for $q>1$)
\begin{equation}\label{S_e}
S_q^{\rm e}(s)=\int_{s}^{\infty}\frac{dx}{1-q}\ln\left[\frac{1}{2^q}\erfc(-x)^q+\frac{1}{2^q}\erfc(x)^q\right],
\end{equation}
where $\erfc(u)=(2/\sqrt{\pi})\int_u^{\infty}e^{-x^2}dx$ is the complementary error function. While $S_q^{\rm e}(s \to -\infty) = \sigma_q$, the constant in the second line of Eq. (\ref{S}), we find that $S_q^{\rm e}(s \to \infty)\approx\frac{q}{q-1}\frac{e^{-s^2}}{4\sqrt{\pi}s^2}$. One can check that $S_q^{\rm e}(s)$ has a well defined limit when $q \to 1$ \cite{Supp_mat}. Similarly, we find that, for large $N$, the cumulants  
$\moy{N_r^p}_c$ of order $p\geq 2$ (see \cite{note_moy} for $p=1$), take the scaling form in the three different regimes (i)-(iii) (see Fig. \ref{Fig_S_var}) 
\be\label{cumul_p}
\moy{N_r^p}_c\approx\begin{cases}
{\cal K}_p^{\rm b}\left(r\right)&\;,\;r = {\cal O}(1)\\
\sqrt{2} \, {\kappa}_p\,r &\;,\;0<r/\sqrt{N} < 1\\
\sqrt{2N} {\cal K}_p^{\rm e}\left(\sqrt{2}(r-\sqrt{N})\right)&\;,\;|\sqrt{N}-r| = {\cal O}(1)\;.
\end{cases}
\ee
where the scaling functions ${\cal K}_p^{\rm b,e}$ as well as the amplitude ${\kappa}_p$ can be again computed explicitly \cite{Supp_mat}, e.g. $\kappa_2 = 1/\sqrt{2\pi}$ (note that $\kappa_p = 0$ for $p$ odd). In particular, in the {\it extended bulk} for $0<r/\sqrt{N} < 1$, we find that all the (even) cumulants are of order ${\cal O}(1)$. Hence in this regime, the fluctuations are not Gaussian. In spite of this non-Gaussianity, the results in the second line of Eqs. (\ref{S}) and (\ref{cumul_p}) show that the entropy $S_q(N,r)$ and $\var{N_r}= \moy{N_r^2}_c$ still have the same $r$-dependence in this regime, and hence are proportional to each other
\begin{equation}
S_q(N,r)\approx\alpha_q\var{N_r} \;, \; \alpha_q = \frac{\sigma_q}{\kappa_2} = \sqrt{2 \pi} S_q^{\rm e}(-\infty) \;, \label{prop}
\end{equation}
for $q>1$. This proportionality is thus similar to Eq. (\ref{S_var}), albeit with a different proportionality constant $\alpha_q \neq (\pi^2/6)(1+1/q)$. In the limit $q \to 1$, we find $\alpha_1 = 3.2017\ldots$ \cite{Supp_mat} (see also \cite{EE_LLL}). Note that this value is in agreement with the lower bound $\alpha_1 \geq 4 \ln 2 = 2.7725\ldots$ valid for non-interacting systems~\cite{Kli06}. This amplitude thus carries the 
signature of the non-Gaussian fluctuations of $N_r$ in the {\it extended bulk}. Finally, at the {\it edge}, the scaling function ${\cal K}_{p}^{\rm e}(s)$ in the third line of Eq.~(\ref{cumul_p}) is given by
\be\label{k_p_e}
{\cal K}_{p}^{\rm e}(s)=-\int_{s}^{\infty}dx \Li_{1-p}\left(-\frac{\erfc(-x)}{\erfc(x)}\right)\;,
\ee
where $\Li_s(x) = \sum_{k\geq 1} x^k/k^s$ is the polylogarithm function \cite{foot_polylog}. While ${\cal K}_{p}^{\rm e}(s \to -\infty) \to \kappa_p$, 
the constant appearing in the second line of Eq.~(\ref{cumul_p}), we show that ${\cal K}_{p}^{\rm e}(s)\approx (-1)^p e^{-s^2}/(4\sqrt{\pi}s^2)$ \cite{Supp_mat}. A comparison of Eqs. (\ref{S_e}) and (\ref{k_p_e}) shows that in the edge regime, the entanglement entropy and the number variance have a rather different $r$-dependence and hence are certainly not proportional to each other (see the inset of Fig. \ref{Fig_S_var}). 
%

%

{\it Entanglement entropy}. The entanglement entropy $S_q(N,r)$ is naturally expressed in terms of the overlap matrix $\mathbb{A}$ in the disk ${\cal D}_r$, where $\mathbb{A}_{kl}=\int_{|z|\leq r}d^2 z \, {\phi^*_k}(z)\phi_l(z)$, as \cite{Kli06} 
\begin{align}\label{S_N}
S_q(N,r)&=\frac{1}{1-q}\Tr[\ln(\mathbb{A}^q+(\mathbb{I}-\mathbb{A})^q)]\\
&=\frac{1}{1-q}\sum_{k=1}^N\ln\left[\lambda_k(r)^q+\left(1-\lambda_k(r)\right)^q\right]\;,\nn
\end{align}
with the $\lambda_k$'s being the eigenvalues of $\mathbb{A}$.
Obtaining the spectrum of $\mathbb{A}$ is in general very hard. However, here, thanks to the spherical symmetry of the system, one can show that $\mathbb{A}$ is diagonal (see also \cite{Kli06}). Indeed, using the expression of $\phi_k(z)$ given in Eq. \eqref{wf}, we obtain (see also \cite{Kli06} in the different context of quantum Hall systems)
\begin{align}
\mathbb{A}_{kl}&=\frac{1}{\sqrt{\Gamma(k)\Gamma(l)}}\int_{0}^r r^{k+l-1}e^{- r^2}dr\int_0^{2\pi}\frac{d\theta}{\pi}e^{-i(k-l)\theta}\nn\\
&=\delta_{kl}\,\lambda_k(r)\;\;{\rm with}\;\;\lambda_k(r)=\frac{\gamma(k,r^2)}{\Gamma(k)}\;,\label{eig}
\end{align}
where the Kronecker delta-function comes from the integration over the angular variable $\theta$ and where $\gamma(a,z)=\int_0^z x^{a-1}e^{-x} dx$ is the lower incomplete gamma function.
The finite $N$ expression of $S_q(N,r)$ is simply obtained by inserting in Eq. \eqref{S_N} the form of $\lambda_k(r)$ given in Eq. \eqref{eig}.
%
%
In the large $N$ limit, we need to analyze the sum in Eq. \eqref{S_N}. As the behavior of $\lambda_k(r)$ is different in each spatial region (i), (ii) and (iii) of Fig. \ref{Fig_S_var}, we will analyze these three cases separately.

(i) In the deep bulk for $r={\cal O}(1)$, from Eq. \eqref{eig} the eigenvalues $\lambda_k(r)=\gamma(k,r^2)/\Gamma(k)$ are independent of $N$. In this regime, the sum over $k$ converges and the scaling function $S_q^{\rm b}(r)$ is obtained by replacing the finite sum in the second line of Eq. \eqref{S_N} by an infinite sum \cite{Supp_mat}. For $r\to 0$, we obtain the asymptotic behavior $S_q^{\rm b}(r)\approx q/(q-1) r^2$, for $q>1$ while $S_1^{\rm b}(r)\approx -2r^2\ln r$. For $r\to \infty$, we obtain $S_q^{\rm b}(r)\approx \sqrt{2}r \sigma_q$ \cite{Supp_mat}, smoothly matching the extended bulk result in the second line of Eq. \eqref{S}. Note that for $q\to 1$, this result is in agreement with the result found in \cite{EE_LLL}, in the context of quantum Hall systems. 

(ii) In the extended bulk, we set $r = \zeta \sqrt{N}$ (with $\zeta < 1$) and study the eigenvalues $\lambda_k(r = \zeta \sqrt{N})$. In this case, the $N$-dependence comes only from the argument $r = \zeta \sqrt{N}$ and in the large $N$ limit, $\lambda_k(r = \zeta \sqrt{N})$ can be estimated via a saddle point method, yielding \cite{gamma_exp}
\be\label{l_eb}
\lambda_k\left(r = \zeta \sqrt{N}\right)\approx F\left(\frac{k-N \zeta^2}{\sqrt{2N}\zeta}\right)\;,\;F(u)=\frac{1}{2}\erfc(u)\;.
\ee
Inserting this scaling form in Eq. \eqref{S_N}, it is then natural to transform the sum over $k$ into a Riemann integral over $x=(k-N \zeta^2)/(\sqrt{2N}\zeta)$. Finally, taking the limit $N\to \infty$ in the bounds of integration, we obtain the result given in the second line of Eq. (\ref{S}). 
Note that in this regime, the area law is verified as $S_q(N,r)\propto r$, in contrast with the case of free \cite{CMV12_2} and trapped fermions \cite{V12,CLM15} in dimension $d$ where there are logarithmic corrections $S_q(N,{\cal D}_l)\propto l^{d-1}\log l$ for a box ${\cal D}_l$ of linear size $l$. This is consistent with Refs. \cite{Area_law_1,Area_law_d,Area_law_d_2} as area laws are expected to hold for systems with correlation functions decreasing exponentially (or faster), which is indeed the case for this system, where the two point-correlation function decays as a Gaussian (see e.g. \cite{LGMS}).

(iii) At the edge, we set $r=\sqrt{N}+s/\sqrt{2}$ with $s={\cal O}(1)$ and we find that the eigenvalues take the scaling form
\be\label{l_e}
\lambda_{N-k}(r)\approx F\left(\sqrt{2}(\sqrt{N}-r)-\frac{k}{\sqrt{2N}}\right)\;,
\ee
where $F(u)$ is given in Eq. (\ref{l_eb}). Inserting this scaling form in Eq. \eqref{S_N}, we now transform the sum over $k$ into a Riemann integral over $x=(k-N)/\sqrt{2N}$, yielding the scaling function given in Eq.~\eqref{S_e}. Note that the limit $r\to \sqrt{N}$ of the extended bulk result in the second line of \eqref{S} matches smoothly with the asymptotic expansion of the edge result $S_q^{\rm e}(s)\approx\sigma_q$ for $s\to -\infty$.

%
%
%
%
{\it Full Counting Statistics}. 
To compute the cumulants of $N_r$, it is useful to study the Centered Cumulant Generating Function (CCGF) $\chi_{\mu}(r)=\ln\moy{e^{-\mu(N_r-\moy{N_r})}}$. In fact, the quantity $\langle e^{-\mu N_r} \rangle$ can be simply expressed in terms of the overlap matrix ${\mathbb A}$ defined above Eq. (\ref{S_N}). After standard manipulations \cite{Supp_mat}, one finds 
\be\label{overlap2}
\langle e^{-\mu N_r} \rangle = \det\left[ {\mathbb I}_N - (1-e^{-\mu}) {\mathbb{A}}\right] \;,
\ee  
where ${\mathbb I}_N$ denotes the $N \times N$ identity matrix. Using the diagonal structure of ${\mathbb A}$ (\ref{eig}), $\chi_\mu(r)$ reads for any finite $N$
\be\label{CCGF}
\chi_\mu(r) = \sum_{k=1}^N \left(\mu\,\lambda_k(r) + \ln\left[1-(1-e^{-\mu})\lambda_k(r) \right] \right) \;,
\ee
in terms of the eigenvalues $\lambda_k(r)$ of ${\mathbb{A}}$ given in (\ref{eig}) and where we have used $\langle N_r \rangle = \sum_{k=1}^N \lambda_k(r)$. 
%
%
After some manipulations \cite{Supp_mat}, we obtain from Eq.~\eqref{CCGF} an exact expression for $\moy{N_r^p}_c$ at any order $p\geq 2$ \cite{foot_ddp}
\be\label{cumul_N}
\moy{N_r^p}_c=(-1)^{p+1}\sum_{k=1}^N \Li_{1-p}\left(1-\frac{1}{\lambda_k(r)}\right)\;.
\ee
Note that for an integer $p$, $\Li_{1-p}(1-1/x)$ is a polynomial in $x$ of degree $p$~\cite{Supp_mat}.
In the large $N$ limit, the behavior of $\moy{N_r^p}_c$ is analysed in a similar manner as $S_q(N,r)$, leading to the scaling forms described in Eq. \eqref{cumul_p}, and in particular with the scaling function at the edge in (\ref{k_p_e}).

In conclusion, we have studied the ground state properties of a system of $N$ 
non-interacting fermions in a $2d$ harmonic trap with frequency $\omega$ in a rotating frame at angular frequency $\Omega$, such that $0 < \omega - \Omega \ll \omega$. We have obtained exact expressions for the R\'enyi entanglement entropy $S_q(N,r)$ as well as for the FCS for a disk of radius $r$, in various scaling regimes for large $N$, where the fermions's density is uniform on the disk of radius $\sqrt{N}$. Far from the edge, $S_q(N,r)$ is proportional to the number variance, which could possibly be measured experimentally, thus providing an indirect access to the entanglement entropy, which is usually hard to measure directly. However, close to the edge, i.e. for $r \approx \sqrt{N}$ we have shown that both observables have instead a quite different $r$-dependence (see Fig. \ref{Fig_S_var}). Our results thus demonstrate that one should be careful when the entanglement entropy is measured indirectly via the number variance, since a relation between both quantities, such as in Eq. (\ref{prop}), actually does not necessarily exist in all regions of space.   

Since the Hamiltonian in the rotating frame (\ref{Ham}) is symmetric under the transformation ${\bf x} \leftrightarrow - {\bf p}$ (in scaled units), our results for the FCS in position space immediately translate to FCS in momentum space, which could then be measured using time-of-flight experiments. This naturally raises the question of the correlations between the positions and momenta of fermions. For instance, it would be very interesting to investigate the Wigner function, as it was recently done for (non-rotating) trapped fermions \cite{wigner}.

%

{\it Acknowledgments:} We would like to thank J. Dalibard for insightful correspondence. This work was partially supported by ANR grant ANR-17-CE30-0027-01 RaMaTraF.

{}

\begin{widetext}

\newpage

\begin{center}
{\Large \bf Supplementary Material for {\it Entanglement Entropy and Full Counting Statistics for $2d$-Rotating Trapped Fermions}}
\end{center}




\maketitle

\section{Asymptotic expansions for the R\'enyi entanglement entropy}
In this section we give the details of the derivations of the R\'enyi entanglement entropy $S_q(N,r)$ together with its asymptotic behaviors in the three different regimes: (i) in the deep bulk, (ii) in the extended bulk and (iii) at the edge.

\subsection{(i) Deep bulk}

In the deep bulk, $\lambda_k(r)$ is independent of $N$ and the asymptotic large $N$ behavior of $S_q(N,r)$ is then obtained simply by replacing the finite sum in Eq. (13) of the main text by an infinite sum. The scaling form valid for $q>1$ reads
\be
S_q(N,r)\approx S_{q}^{\rm b}(r)=\frac{1}{1-q}\sum_{k=1}^{\infty}\ln\left[\left(\frac{\gamma(k,r^2)}{\Gamma(k)}\right)^q+\left(\frac{\Gamma(k,r^2)}{\Gamma(k)}\right)^q\right]\;,\label{S_b}
\ee
while in the case $q\to 1$, it reads
\be
S_{q=1}(N,r) = S_{\rm vN}(N,r)\approx S_{\rm vN}^{\rm b}(r)=-\sum_{k=1}^{\infty}\frac{\gamma(k,r^2)}{\Gamma(k)}\ln\left[\frac{\gamma(k,r^2)}{\Gamma(k)}\right]-\sum_{k=1}^{\infty}\frac{\Gamma(k,r^2)}{\Gamma(k)}\ln\left[\frac{\Gamma(k,r^2)}{\Gamma(k)}\right]\;.\label{S_vN_b}
\ee
In the limit of small interval, $r\to 0$, one can use the small $x$ behaviour of $\gamma(k,x)$,
\be
\frac{\gamma(k,x)}{\Gamma(k)}=\frac{1}{\Gamma(k)}\int_0^{x}t^{k-1}e^{-t}dt=\left[\frac{t^{k}}{\Gamma(k+1)}e^{-t}\right]_0^x+\frac{1}{\Gamma(k+1)}\int_0^{x}t^{k}e^{-t}dt\approx \frac{x^k}{\Gamma(k+1)}+ O(x^{k+1})\;,
\ee
and only retain the term of lowest order $k=1$ in Eqs. \eqref{S_b} and \eqref{S_vN_b}. In this limit, the scaling function reads
\be\label{S_b_-}
 S_{q}^{\rm b}(r)\approx \frac{q}{q-1}r^2\;\;{\rm and}\;\;S_{\rm vN}^{\rm b}(r)\approx -2r^2\ln r\;,\;\;{\rm for}\;\;r\to 0\;.
\ee
On the other hand, in the limit of large interval $r\to \infty$ and in the regime $|k-r^2|\sim r$, the eigenvalues $\lambda_k(r)$ take the scaling form \cite{gamma_exp} (this form is obtained by a saddle-point approximation, see for instance \cite{LGMS}),
\be\label{large_s_scal}
\lambda_k(r)=\frac{\gamma(k,r^2)}{\Gamma(k)}\approx \frac{1}{2}\erfc\left(\frac{k-r^2}{\sqrt{2}r}\right)\;.
\ee
We replace in Eqs. \eqref{S_b} and \eqref{S_vN_b} the discrete sums over $k$ by an integral over $x=(k-r^2)/(\sqrt{2}r)$. In the limit of large $r$, the lower bound of integration $-r/\sqrt{2}$ is replaced by $-\infty$. This yields
\begin{align}
&S_{q}^{\rm b}(r)\approx \sqrt{2}\,\sigma_q \, r\;\;\;{\rm and}\;\;S_{\rm vN}^{\rm b}(r)\approx \sqrt{2}\,\sigma_1 \, r\;,\;\;{\rm for}\;\;r\to \infty\;,\\
&{\rm where}\;\;\sigma_q=\int_{-\infty}^{\infty}\frac{dx}{1-q}\ln\left[\frac{1}{2^q}\erfc(x)^q+\frac{1}{2^q}\erfc(-x)^q\right]\;\;{\rm and}\;\;\sigma_1=-\int_{-\infty}^{\infty}dx\erfc(x)\ln\left[\frac{1}{2}\erfc(x)\right] = 1.2773\ldots\;. \label{sigma_q}
\end{align}
This result matches smoothly with the extended bulk result (see below). Note that the result for $\sigma_1$ can be recovered by expanding the expression of $\sigma_q$ for $q \to 1$.

\subsection{(ii) Extended bulk}
In the extended bulk, inserting in Eq. (13) of the main text the scaling form for $\lambda_k(r)$ in Eq. (15) of the main text, we replace the discrete sum over $k$ by an integral over $x=(k-N\zeta^2)/(\sqrt{2N}\zeta)$ with $0<\zeta=r/\sqrt{N}<1$. It reads
\be
S_q(N,r)\approx \sqrt{2}\,r\int_{-\sqrt{N/2}\zeta^{-1}}^{\sqrt{N/2}(\zeta^{-1}-\zeta)}\frac{dx}{1-q}\ln\left[\frac{1}{2^q}\erfc(x)^q+\frac{1}{2^q}\erfc(x)^q\right]\;.
\ee
As $0<\zeta<1$, we replace in the large $N$ limit the bounds of integration by $\pm \infty$, and we obtain that $S_q(N,r) \approx \sqrt{2} \sigma_q\, r$ where $\sigma_q$ is given above in Eq. (\ref{sigma_q}), as announced in the second line of Eq. (8) in the main text. For $q=1$, the result is similar and reads
\be
S_{\rm vN}(N,r)\approx -\sqrt{2}\,r\int_{-\infty}^{\infty}dx\erfc(x)\ln\left[\frac{1}{2}\erfc(x)\right]=\sqrt{2}\sigma_1 \, r=\frac{\alpha_1 r}{\sqrt{\pi}}\;,
\ee
where $\sigma_1$ is given in (\ref{sigma_q}).

\subsection{(iii) Edge regime}

At the edge, inserting in Eq. (13) of the main text the scaling form of Eq. (16) of the main text, we replace the discrete sum over $k$ by an integral over $u=(k-N)/\sqrt{2N}$. For fixed $s=\sqrt{2}(r-\sqrt{N})=O(1)$, it reads
\be
S_q(N,r)\approx \sqrt{2N}\int_{-\sqrt{N/2}}^{0}\frac{du}{1-q}\ln\left[\frac{1}{2^q}\erfc(-s+u)^q+\frac{1}{2^q}\erfc(s-u)^q\right]\;.
\ee
Making an additional change of variable $u\to x=s-u$ and taking the limit in the new upper bound of the integral $\sqrt{N/2}+s\to+ \infty$, this yields the expression (for $q>1$)
\be
S_q(N,r)\approx \sqrt{2N} S_{q}^{\rm e}(\sqrt{2}(r-\sqrt{N}))\;,\;\;{\rm with}\;\; S_{q}^{\rm e}(s)=\int_{s}^{\infty}\frac{dx}{1-q}\ln\left[\frac{1}{2^q}\erfc(x)^q+\frac{1}{2^q}\erfc(-x)^q\right]\;,\label{S_e}
\ee
as announced in Eqs. (8) and (9) of the main text. For $q=1$, the expression reads
\be
S_{\rm vN}(N,r)\approx \sqrt{2N} S_{\rm vN}^{\rm e}(\sqrt{2}(r-\sqrt{N}))\;,\;\;{\rm with}\;\;S_{\rm vN}^{\rm e}(s)=-\int_{s}^{\infty}\frac{dx}{2}\left(\erfc(x)\ln\left[\frac{1}{2}\erfc(x)\right]+\erfc(-x)\ln\left[\frac{1}{2}\erfc(-x)\right]\right)\;.\label{S_vN_e}
\ee

As the integrals in Eqs. \eqref{S_e} and \eqref{S_vN_e} converge for $s\to -\infty$, it is straightforward to obtain that $S_{q}^{\rm e}(s \to -\infty)\to \sigma_q$ for $q\geq 1$, where $\sigma_q$ is given in Eq. (\ref{sigma_q}). On the other hand, in the limit $s\to +\infty$, using the asymptotic expansion $\erfc(-x)\approx 2-e^{-x^2}/(\sqrt{\pi}x)$ for $x\to \infty$ in Eq. \eqref{S_e}, we obtain at leading order
\be\label{S_e_+}
S_{q}^{\rm e}(s)\approx \frac{q}{q-1}\frac{e^{-s^2}}{4\sqrt{\pi}s^2}\;\;{\rm and}\;\;S_{\rm vN}^{\rm e}(s)\approx \frac{e^{-s^2}}{4\sqrt{\pi}}\;,\;\;{\rm for}\;\;s\to +\infty\;,
\ee
as announced below Eq. (9) of the main text.

\section{Full Counting Statistics for finite $N$}
In this section, we derive the expression of the cumulants given in Eq. (19) of the main text. First we obtain the finite $N$ expression of the Centred Cumulant Generating Function (CCGF) given in Eq. (18) of the main text, and then we extract the cumulants.
We first compute the moment generating function, defined as the Laplace transform of the PDF of $n_r=N_r-\moy{N_r}$,
\be\label{MGF_1}
\moy{e^{-\mu(N_r-\moy{N_r})}}=e^{\mu\moy{N_r}}\moy{e^{-\mu N_r}}=e^{\mu\moy{N_r}}\int d^2 z_1 \cdots \int d^2z_N e^{-\mu \sum_{k=1}^N I_r(z_k)}|\Psi_0(z_1,\cdots,z_N)|^2\;,
\ee 
where we have introduced the indicator function $I_r(z)=1$ if $r\in {\cal D}_r$ and $I_r(z)=0$ otherwise. Using that the many-body wave function is a Slater determinant $\Psi_0(z_1, \cdots, z_N) = \det_{1\leq j,k \leq N}\phi_k(z_j)/\sqrt{N!}$ and together with the Cauchy-Binet identity, Eq. \eqref{MGF_1} can be expressed as
\be\label{MGF_2}
\moy{e^{-\mu(N_r-\moy{N_r})}}=e^{\mu\moy{N_r}}\det_{1\leq k,l\leq N}\left[\int d^2 z e^{-\mu I_r(z)}{\phi^*_k}(z)\phi_l(z)\right]\;.
\ee
Next, we use that $e^{-\mu I_r(z)}=1-(1-e^{-\mu})I_r(z)$, since $I_r(z) = 0$ or $I_r(z)=1$. Using the orthonormality of the wave functions (see Eq. (14) in the main text) in Eq. \eqref{MGF_2}, we obtain
\be\label{MGF_3}
\moy{e^{-\mu(N_r-\moy{N_r})}}=e^{\mu\moy{N_r}}\det_{1\leq k,l\leq N}\left[\delta_{k,l}-(1-e^{-\mu})\int_{|z|\leq r} d^2 z\,{\phi^*_k}(z)\phi_l(z)\right]=e^{\mu\moy{N_r}}\det_{1\leq k,l\leq N}\left[\delta_{k,l}-(1-e^{-\mu})\mathbb{A}_{k,l}\right]\;,
\ee
where we recall that $\mathbb{A}$ is the overlap matrix. Note that at this stage the result is valid for any system of non-interacting fermions. In this case, the overlap matrix is diagonal $\mathbb{A}_{k,l}=\delta_{k,l}\lambda_k(r)$ (see Eq. (14) in the main text) and the determinant in Eq. \eqref{MGF_3} reduces to a finite product of $N$ terms. Taking the logarithm, we finally obtain the CCGF
\be\label{CCGF}
\chi_\mu(r)=\ln\moy{e^{-\mu(N_r-\moy{N_r})}}=\sum_{k=1}^N \ln[1-(1-e^{-\mu})\lambda_k(r)]+\sum_{k=1}^N \mu \lambda_k(r)\;,
\ee
where we used $\moy{N_r}=\sum_{k=1}^N \lambda_k(r)$.
From Eq. \eqref{CCGF}, we factorise $(1-\lambda_k(r))$ in the argument of the logarithm to obtain
\be\label{CCGF_2}
\chi_\mu(r)= \sum_{k=1}^N\ln[1-\lambda_k(r)]+\mu\moy{N_r}+\sum_{k=1}^N\ln\left[1+e^{-\mu}\frac{\lambda_k(r)}{1-\lambda_k(r)}\right]\;.
\ee
Using next the Taylor series $\ln(1+x)=\sum_{l=1}^\infty (-1)^{l+1}x^l/l$, this yields
\be\label{CCGF_3}
\chi_\mu(r)= \sum_{k=1}^N\ln[1-\lambda_k(r)]+\mu\moy{N_r}+\sum_{l=1}^{\infty}\sum_{k=1}^N (-1)^{l+1} \frac{e^{-l\mu}}{l}\left(\frac{\lambda_k(r)}{1-\lambda_k(r)}\right)^l\;.
\ee
To obtain the expression of the cumulants, we compare this equation with the cumulant expansion of the CCGF
\be\label{CCGF_cumul}
\chi_\mu(r)= \sum_{p=2}^\infty \frac{(-1)^p}{p!}\moy{N_r^p}_c \, \mu^p\;.
\ee
Identifying the coefficient in $\mu^p$ in Eqs. \eqref{CCGF_3} and \eqref{CCGF_cumul}, we obtain the identity valid for all $p\geq 2$,
\be\label{cumul_1}
\moy{N_r^p}_c=\sum_{k=1}^N \sum_{l=1}^{\infty} (-1)^{l+1} l^{p-1}\left(\frac{\lambda_k(r)}{1-\lambda_k(r)}\right)^l=-\sum_{k=1}^N \Li_{1-p}\left(-\frac{\lambda_k(r)}{1-\lambda_k(r)}\right)\;,
\ee
where we used that $\Li_s(x)=\sum_{k=1}^\infty k^{-s} x^k$ is the polylogarithm function.

Note that if instead, we factorise $\lambda_k(r)e^{-\mu}$ in the argument of the logarithm in Eq. \eqref{CCGF} and use again the Taylor series $\ln(1+x)=\sum_{l=1}^\infty (-1)^{l+1}x^l/l$, we obtain a different expression for the CCGF
\begin{align}\label{CCGF_4}
\chi_\mu(r)&= -\mu\sum_{k=1}^N\ln[\lambda_k(r)]+\mu\moy{N_r}+\sum_{l=1}^{\infty}\ln\left[1+\frac{1-\lambda_k(r)}{\lambda_k(r)}e^{\mu}\right]\nn\\
&=-\mu\sum_{k=1}^N\ln[\lambda_k(r)]+\mu\moy{N_r}+\sum_{k=1}^N \sum_{l=1}^\infty(-1)^{l+1} \frac{e^{l\mu}}{l}\left(\frac{1-\lambda_k(r)}{\lambda_k(r)}\right)^l\;.
\end{align}
Identifying now the coefficient in $\mu^p$ in Eqs. \eqref{CCGF_4} and \eqref{CCGF_cumul}, we obtain an additional {identity} valid for all $p\geq 2$,
\be\label{cumul_2}
\moy{N_r^p}_c=\sum_{k=1}^N \sum_{l=1}^{\infty} (-1)^{l+p+1} l^{p-1}\left(\frac{1-\lambda_k(r)}{\lambda_k(r)}\right)^l=(-1)^{p+1}\sum_{k=1}^N \Li_{1-p}\left(-\frac{1-\lambda_k(r)}{\lambda_k(r)}\right)=(-1)^{p+1}\sum_{k=1}^N \Li_{1-p}\left(1-\frac{1}{\lambda_k(r)}\right)\;,
\ee
which is the formula given in Eq. (20) of the main text.
From Eqs. \eqref{cumul_1} and \eqref{cumul_2} we obtain immediately the well know property \cite{erdelyi}
\be\label{Li_inverse}
\Li_{1-l}\left(1-\frac{1}{x}\right)=\Li_{1-l}\left(\frac{x-1}{x}\right)=(-1)^l \Li_{1-l}\left(\frac{x}{x-1}\right)=(-1)^l \Li_{1-l}\left(1-\frac{1}{1-x}\right)\;,
\ee
which will be useful for later purpose. Additionally, using the definition of the polylogarithm, we obtain the small $x$ behaviour
\be\label{Li_small}
\Li_{1-l}\left(1-\frac{1}{x}\right)=(-1)^l \Li_{1-l}\left(-\frac{x}{1-x}\right)\approx (-1)^{l+1}x\;, {\rm for}\;\;x\to 0\;,
\ee
which will be useful to obtain asymptotic expansions in the next section.
%
%
%
%

Note finally that the function $\Li_{1-p}(1-1/x)$ is a polynomial function of degree $p$ in the variable $x$. It can be shown using the identities
\be
\frac{e^{-z}}{1-e^{-z}}=\sum_{l=1}^{\infty}(e^{-z})^l\;\;{\rm and}\;\;\frac{-e^{-z}}{1+e^{-z}}=\sum_{l=1}^{\infty}(-e^{-z})^l\;,
\ee
and deriving $p-1$ time these expressions with respect to $z$, to obtain
\begin{align}\label{Q_P}
\frac{P_{p}(e^{-z})}{(1-e^{-z})^p}&=\sum_{l=1}^{\infty}(-l)^{p-1}(e^{-z})^l=(-1)^{p-1}\Li_{1-p}\left(e^{-z}\right)\\
\frac{Q_{p}(e^{-z})}{(1+e^{-z})^p}&=\sum_{l=1}^{\infty}(-l)^{p-1}(-e^{-z})^l=(-1)^{p-1}\Li_{1-p}\left(-e^{-z}\right)\;,\nn
\end{align}
where $P_p(y)$ and $Q_p(y)$ are polynomial functions of degree $p$ in the variable $y$. Finally, replacing $e^{-z}=1/x-1<1$ if $1>x>1/2$ in the first line of Eq. \eqref{Q_P} or $e^{-z}=x/(1-x)<1$ if $0<x<1/2$, we obtain
\begin{align}
&(-1)^{p-1}\Li_{1-p}\left(1-\frac{1}{x}\right)=x^p P_{p}\left(-\frac{1-x}{x}\right)=\tilde P_p(x)\;,\nn\\
&(-1)^{p-1}\Li_{1-p}\left(1-\frac{1}{1-x}\right)=(1-x)^p Q_p\left(-\frac{x}{1-x}\right)=\tilde Q_p(x)
\end{align}
where $\tilde P_p(x)$ and $\tilde Q_p(x)$ are polynomial functions of degree $p$ in the variable $x$.

\section{Asymptotic results for the cumulants in the large $N$ limit} 
In this section we analyze the behavior of the cumulants $\moy{N_r^p}_c$ in each of the three spatial regions. We also provide asymptotic expansions of these cumulants in the deep bulk (i) and at the edge (iii).

\subsection{(i) Deep bulk}

In the deep bulk, the eigenvalues $\lambda_k(r)$ do not depend on $N$, at leading order for large $N$. Therefore, in the large $N$ limit, the cumulant scaling function is obtained by replacing the finite sum over $k$ in Eq. (19) of the main text by an infinite sum. This yields
\be
\moy{N_r^p}_c\approx {\cal K}^{\rm b}_p(r)\;,\;\;{\cal K}^{\rm b}_p(r)=(-1)^{p+1}\sum_{k=1}^{\infty}\Li_{1-p}\left(-\frac{\Gamma(k,r^2)}{\gamma(k,r^2)}\right)\;. \label{cumul_b}
\ee
In the small $r$ limit, the eigenvalues are small $\lambda_k(r)\sim r^{2k}/\Gamma(k+1)\ll 1$ and we may retain only the term for $k=1$. Using additionally the expansion of Eq. \eqref{Li_small}, we obtain
\be
{\cal K}^{\rm b}_p(r)\approx r^2\;,\;\;{\rm for}\;\;r\to 0\;,
\ee
independently of $p$. Hence all the cumulants are equal ($\moy{N_r^p}\approx r^2$ for $r \ll 1$ and all $p\geq 1$), which is characteristic of a Poisson distribution. Note that the mean number of fermions inside the disk of radius $r$ can be simply obtained from the fermion density  
\be
\langle N_r\rangle=2\pi\int_{0}^{r}r'\rho_N(r')dr'\approx r^2\;,\;\;{\rm for}\;\;r = {\cal O}(1) \;\; {\rm as} \;\; N \to \infty\;,
\ee
Hence in this deep bulk regime $r \to 0$, the distribution of $N_r$ is Poissonian with parameter $r^2$. Additionally, we note that, in this bulk regime, the R\'enyi entropy $S_q^{\rm b}(r)$ also behaves quadratically with $r$ (as discussed in the main text)
\begin{eqnarray}\label{S_small_r}
S_q^{\rm b}(r) \approx \frac{q}{q-1}\,r^2 \;.
\end{eqnarray}
Hence, in this bulk regime, the proportionality between the number variance and the entanglement entropy is restored, i.e. 
\begin{eqnarray}\label{rel_deep_bulk}
S_q^{\rm b}(r) \approx\frac{q}{q-1} {\rm Var}(N_r)  \;.
\end{eqnarray}
Thus the proportionality constant $q/(q-1)$ in this bulk regime is different from $\alpha_q$ in the extended bulk regime, as given in Eq. (11) in the main text.

For a large disk $r\to +\infty$, we use the scaling form in Eq. \eqref{large_s_scal} and rewrite the sum in Eq. \eqref{cumul_b} as an integral over $x=(k-r^2)/(\sqrt{2}r)$. It reads
\be
{\cal K}^{\rm b}_p(r)\approx (-1)^{p+1}\sqrt{2}r\int_{-r/\sqrt{2}}^{\infty}dx \Li_{1-p}\left(-\frac{\erfc(-x)}{\erfc(x)}\right) \;.
\ee
As $r\to \infty$, we replace the lower bound of integration by $-\infty$. This yields
\be
{\cal K}^{\rm b}_p(r)\approx\sqrt{2}r \kappa_p \;,\;\;{\rm for}\;\;r\to +\infty\;\;{\rm where}\;\; \kappa_p=(-1)^{p+1}\int_{-\infty}^{\infty}dx \Li_{1-p}\left(-\frac{\erfc(-x)}{\erfc(x)}\right)\;.
\ee
This expression matches smoothly with the extended bulk result (see below).

\subsection{(ii) Extended bulk}

In the extended bulk, we insert in Eq. (19) of the main text the scaling form for the eigenvalues in Eq. (15) of the main text. Replacing the sum over $k$ by a Riemann integral over $x=(k-N\zeta^2)/(\sqrt{2}\zeta)$ and taking the limit $N\to \infty$ in the bounds of integration, we obtain
\be\label{k_p_1}
\moy{N_r^p}_c\approx \sqrt{2}r\kappa_p\;\;{\rm with}\;\;\kappa_p=(-1)^{p+1}\int_{-\infty}^{\infty}dx \Li_{1-p}\left(-\frac{\erfc(-x)}{\erfc(x)}\right)\;.
\ee
Changing the integration variable $x\to -x$, the expression of $\kappa_p$ reads
\be\label{k_p_2}
\kappa_p=(-1)^{p+1}\int_{-\infty}^{\infty}dx \Li_{1-p}\left(-\frac{\erfc(x)}{\erfc(-x)}\right)\;.
\ee
Using the property in Eq. \eqref{Li_inverse}, we obtain the simplified expression 
\be\label{k_p_3}
\kappa_p=-\int_{-\infty}^{\infty}dx \Li_{1-p}\left(-\frac{\erfc(-x)}{\erfc(x)}\right)\;.
\ee
By comparing Eqs. \eqref{k_p_1} and \eqref{k_p_3}, we realize that the odd cumulants vanish in the extended bulk $\kappa_{2p+1}=-\kappa_{2p+1}=0$.

The coefficient $\kappa_2$ for the variance can be obtained explicitly as follows. Using Eq. (\ref{k_p_3}) specialised to $p=2$ and using $\Li_{-1}(-x)=-x/(1+x)^2$, we obtain
\be\label{k2_1}
\kappa_2=\int_{-\infty}^{\infty}\frac{\erfc(-x)dx}{\erfc(x)\left(1+\frac{\erfc(-x)}{\erfc(x)}\right)^2}=\int_{-\infty}^{\infty}\frac{dx}{4}\erfc(-x)\erfc(x)\;,
\ee
where we used $\erfc(x)+\erfc(-x)=2$.
Using an integration by part and the identity $(\erfc(x)\erfc(-x))'=2[\erfc(x)-\erfc(-x)]e^{-x^2}/\sqrt{\pi}=-4e^{-x^2}\erf(x)/\sqrt{\pi}$, we express Eq. \eqref{k2_1} as
\be
\kappa_2=\left[\frac{x}{4}\erfc(x)\erfc(-x)\right]_{-\infty}^{\infty}+\int_{-\infty}^{\infty}\frac{x}{\sqrt{\pi}}e^{-x^2}\erf(x)dx=\int_{-\infty}^{\infty}\frac{x}{\sqrt{\pi}}e^{-x^2}\erf(x)dx\;.
\ee
Finally, making an additional integration by parts, we obtain the final result
\be\label{k2}
\kappa_2=\left[-\frac{e^{-x^2}}{2\pi}\erf(x)\right]_{-\infty}^{\infty}+\frac{2}{\pi}\int_{-\infty}^{\infty}e^{-2x^2}dx=\frac{1}{\sqrt{2\pi}}\;,
\ee
as announced in the text below Eq. (10). In particular, using Eq. \eqref{k2}, we express the variance in the extended bulk as ${\rm Var}{(N_r)}\approx r/\sqrt{\pi}$.
\subsection{(iii) Edge regime}

\begin{figure}[h]
\centering
\includegraphics[width=0.5\textwidth]{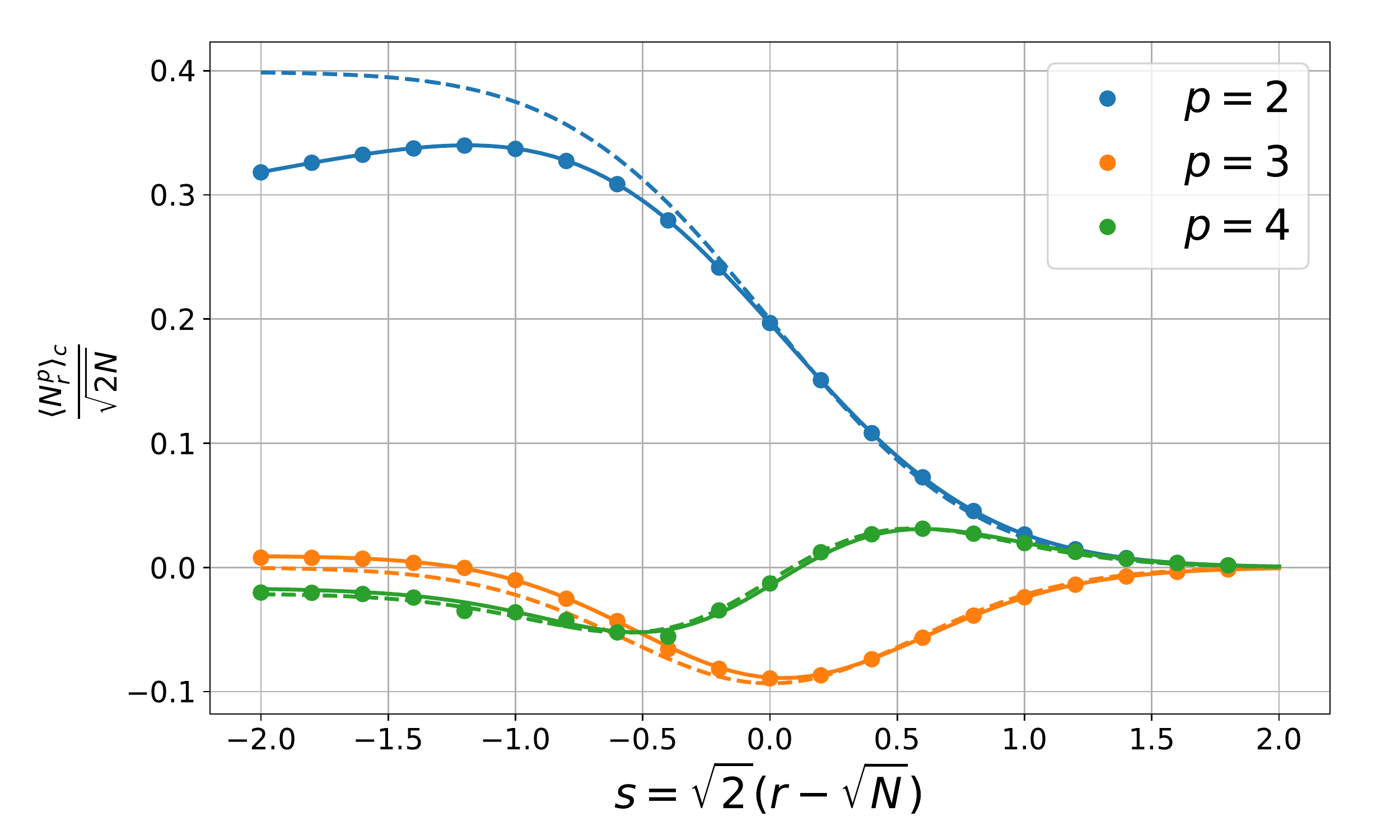}
\caption{Plot of the rescaled cumulants $\moy{N_r^p}_c/\sqrt{2N}$ for $p=2$ (blue dots), $p=3$ (orange dots) and $p=4$ (green dots) as a function of the rescaled position at the edge $s=\sqrt{2}(r-\sqrt{N})$ obtained numerically by diagonalization of $1.2\times 10^6$ complex Ginibre matrices of linear size $N=50$. They are compared with the analytical prediction for finite $N$ (solid lines) and the asymptotic result for $N\to \infty$ (dashed lines).}\label{Fig_cumul}
\end{figure}

At the edge, we insert in Eq. (19) of the main text the scaling form for $\lambda_k(r)$ in Eq. (16) of the main text. Replacing the sum over $k$ by an integral over $u=(k-N)/\sqrt{2N}$ with fixed $s=\sqrt{2}(r-\sqrt{N})$, this yields
\be
\moy{N_r^p}_c\approx \sqrt{2N}(-1)^{p+1} \int_{-\sqrt{N/2}}^{0}du \Li_{1-p}\left(-\frac{\erfc(-u+s)}{\erfc(u-s)}\right)\;.
\ee
Making an additional change of variable $u\to x=s-u$, using Eq. \eqref{Li_inverse} and taking the upper bound of integration $\sqrt{N/2}+s\to \infty$, we obtain
\be\label{psi_e}
\moy{N_r^p}_c\approx \sqrt{2N}{\cal K}_p^{\rm e}(\sqrt{2}(r-\sqrt{N}))\;\;{\rm with}\;\;{\cal K}_p^{\rm e}(s)=-\int_{s}^{\infty}dx \Li_{1-p}\left(-\frac{\erfc(-x)}{\erfc(x)}\right)\;.
\ee 
In Fig. \ref{Fig_cumul}, we show a plot of these functions ${\cal K}_p^{\rm e}(s)$ for $p=2,3,4$ and compare them to numerical simulations. Note that the discrepancy between the analytical prediction and the numerics observed for large negative $s$ is a finite $N$ effect. The integral in Eq. \eqref{psi_e} is convergent for $s\to -\infty$ such that we obtain straightforwardly ${\cal K}_p^{\rm e}(s \to -\infty) =  \kappa_p$ in this limit. 
For $s\to +\infty$, the argument of $\Li_{1-p}(1-1/u)$ is small $u=\erfc(x)/2\approx e^{-x^2}/(2\sqrt{\pi}x)$ and from Eq. \eqref{Li_small}, $-\Li_{1-p}(1-1/u)\approx (-1)^{p} u$. Inserting this expression in Eq. \eqref{psi_e}, we obtain for $p\geq 2$,
\be\label{psi_e_+}
{\cal K}_p^{\rm e}(s)\approx (-1)^p\frac{e^{-s^2}}{4\sqrt{\pi}s^2}\;,\;\;{\rm for}\;\;s\to +\infty\;.
\ee
To understand this expression, we consider the average number of fermion outside the disk of radius $r=\sqrt{N}+s/\sqrt{2}$, 
\be
\langle \overline{N}_r\rangle=2\pi\int_{r}^{\infty}r'\rho_N(r')dr'\approx\sqrt{N}\int_s^{\infty}\frac{dx}{\sqrt{2}}\erfc(x)\approx\sqrt{2N}\frac{e^{-s^2}}{4\sqrt{\pi}s^2}\;,\;\;{\rm for}\;\;s\to +\infty\;,
\ee
where we used the scaling form for the density at the edge \cite{Forrester_book} $\rho_N(r)=\erfc(s)/(2\pi)$ with $s=\sqrt{2}(r-\sqrt{N})$. The distribution of $\overline{N_r}=N-N_r$ will  be Poissonian with parameter $\langle \overline{N}_r\rangle=\sqrt{2N}\frac{e^{-s^2}}{4\sqrt{\pi}s^2}$, and therefore the cumulants for $p\geq 2$ are given by $\moy{N_r^p}_c=(-1)^p\moy{\overline{N_r}^p}_c=(-1)^p \overline{N_r}$.

From Eq. \eqref{psi_e}, we can obtain explicit expressions for the cumulants at the edge
\begin{align}
{\cal K}_2^{\rm e}(s)&=\int_{s}^{\infty}\frac{dx}{4}\erfc(x)\erfc(-x)=\frac{e^{-s^2}}{2\sqrt{\pi}}\erf(s)+\frac{s}{4}\erfc(s)\erfc(-s)+\frac{1}{2\sqrt{2\pi}}\erfc(\sqrt{2}s)\;,\\
{\cal K}_3^{\rm e}(s)&=\int_s^{\infty}\frac{dx}{4}\left[\erf^3(x)-\erf(x)\right]\;,\;\;{\cal K}_4^{\rm e}(s)=\int_s^{\infty}\frac{dx}{8}\left[-1+4\erf(x)^2-3\erf(x)^4\right]\;.
\end{align}
More generally, ${\cal K}_p^{\rm e}(s)$ is given as an integral of a polynomial of degree $p$ in the variable $\erf(x)$.
Note that we could also have included in these cumulants the leading correction to the mean number $\moy{N_r}$ that takes a similar scaling form 
\be
 \moy{N_r}-N\approx\sqrt{2N}{\cal K}_1^{\rm e}(\sqrt{2}(r-\sqrt{N}))\;,\;\;{\rm with}\;\; {\cal K}_1^{\rm e}(s)=-\int_{s}^{\infty}\frac{dx}{2}\erfc(x)=\frac{s}{2}\erfc(s)-\frac{e^{-s^2}}{2\sqrt{\pi}}\;.
\ee

{}

\end{widetext}


\begin{thebibliography}{}

\bibitem{BDZ08}
I. Bloch, J. Dalibard, W. Zwerger, {\it  Many-body physics with ultracold gases}, Rev. Mod. Phys. {\bf 80}, 885 (2008).

\bibitem{Fermicro1}
L.~W. Cheuk, M.~A. Nichols, M. Okan, T. Gersdorf, R.Vinay, W. Bakr, T. Lompe, M. Zwierlein, {\it  Quantum-gas microscope for fermionic atoms}, Phys. Rev. Lett. {\bf 114},
193001, (2015).

\bibitem{Fermicro2}
 E. Haller, J. Hudson, A. Kelly, D.~A. Cotta, B. Peaudecerf, G.~D. Bruce, S. Kuhr, {\it  Single-atom imaging of fermions in a quantum-gas microscope}, Nat. Phys. {\bf 11}, 738 (2015).

\bibitem{Fermicro3}
 M.~F. Parsons, F. Huber, A. Mazurenko, C.~S. Chiu, W. Setiawan, K. Wooley-Brown, S. Blatt, M. Greiner, {\it  Site-resolved imaging of fermionic $^6$Li in an optical lattice}, Phys. Rev. Lett. {\bf 114}, 213002 (2015).



\bibitem{Lev96}
L.~S. Levitov, H.~W. Lee, G.~B. Lesovik, {\it  Electron counting statistics and coherent states of electric current}, J. Math. Phys. {\bf 37},  4845 (1996).

\bibitem{Been06}
C.~W. Groth, B. Michaelis, C.~W.~J. Beenakker, {\it  Counting statistics of coherent population trapping in quantum dots}, Phys. Rev. B {\bf 74}(12), 125315 (2006).

\bibitem{Gus06}
S. Gustavsson, R. Leturcq, B. Simovi\v{c}, R. Schleser, T. Ihn, P. Studerus, K. Ensslin, D.~C. Driscoll, A.~C. Gossard, {\it  Counting statistics of single electron transport in a quantum dot}, Phys. Rev. Lett. {\bf 96}, 076605 (2006).

\bibitem{Nat08}
S. Hofferberth, I. Lesanovsky, T. Schumm, A. Imambekov, V. Gritsev, E. Demler, J. Schmiedmayer, {\it  Probing quantum and thermal noise in an interacting many-body system}, Nat. Phys. {\bf 4}, 489 (2008).

\bibitem{SRev}
L. Amico,  R. Fazio,  A. Osterloh, V.  Vedral, {\it  Entanglement in many-body systems},  Rev. Mod. Phys. {\bf 80}, 517 (2008).


\bibitem{CC04}
P. Calabrese, J. Cardy, {\it  Entanglement entropy and quantum field theory}, J. Stat. Mech. P06002 (2004).
%

\bibitem{Dubail}
J. Dubail, J.-M. St\'ephan, J. Viti, P. Calabrese,  {\it Conformal Field Theory for Inhomogeneous One-dimensional Quantum Systems: the Example of Non-Interacting Fermi Gases}, SciPost Phys. {\bf 2}, 002 (2017). 



\bibitem{RM04}
G. Refael, J. E. Moore, {\it Entanglement entropy of random quantum critical points in one dimension}, Phys. Rev. Lett. {\bf 93}, 260602 (2004).


\bibitem{MFS09}
M. A. Metlitski, C. A. Fuertes, S. Sachdev, {\it  Entanglement entropy in the $O(N)$ model}, Phys. Rev. B {\bf 80}, 115122  (2009). 


\bibitem{DMH17}
T. Devakul, S. N. Majumdar, D. A. Huse, {\it Probability distribution of the entanglement across a cut at an infinite-randomness fixed point}, Phys. Rev. B {\bf 95}, 104204 (2017).




\bibitem{Kli06}
I. Klich, {\it Lower entropy bounds and particle number fluctuations in a Fermi sea}, J. Phys. A: Math. Gen. {\bf 39}, L85 (2006). 



\bibitem{KL09}
I. Klich, L. Levitov, {\it  Quantum noise as an entanglement meter}, Phys. Rev. Lett. {\bf 102}, 100502 (2009).

\bibitem{Hur11}
H.~F. Song, C. Flindt, S. Rachel, I. Klich, K. Le Hur, {\it  Entanglement entropy from charge statistics: Exact relations for noninteracting many-body systems}, Phys. Rev. B {\bf 83}, 161408(R) (2011).

\bibitem{CLM15}
P. Calabrese, P. Le Doussal, S.~N. Majumdar, {\it  Random matrices and entanglement entropy of trapped Fermi gases}, Phys. Rev. A {\bf 91}(1), 012303 (2015).

\bibitem{V12}
E. Vicari, {\it  Entanglement and particle correlations of Fermi gases in harmonic traps}, Phys. Rev. A {\bf 85}, 062104 (2012).

\bibitem{CMV12_1}
P. Calabrese, M. Mintchev, E. Vicari, {\it  Exact relations between particle fluctuations and entanglement in Fermi gases}, Europhys. Lett. {\bf 98}, 20003 (2012).






\bibitem{CMV11}
P. Calabrese, M. Mintchev, E. Vicari, {\it  Entanglement entropy of one-dimensional gases}, Phys. Rev. Lett. {\bf 107}, 020601 (2011).

\bibitem{us_PRL}
D. S. Dean, P. Le Doussal, S. N. Majumdar, G. Schehr, {\it Finite temperature free fermions and the Kardar-Parisi-Zhang equation at finite time}, 
Phys. Rev. Lett. {\bf 114}, 110402 (2015).




\bibitem{us_EPL}
D. S. Dean, P. Le Doussal, S. N. Majumdar, G. Schehr, {\it Universal ground state properties of free fermions in a d-dimensional trap}, 
Europhys. Lett. {\bf 112}, 60001 (2015).


\bibitem{us_PRA}
D. S. Dean, P. Le Doussal, S. N. Majumdar, G. Schehr, {\it Non-interacting fermions at finite temperature in a d-dimensional trap: universal correlations}, 
Phys. Rev. A {\bf 94}, 063622 (2016).




\bibitem{Ricardo}
R. Marino,  S.~N. Majumdar,  G. Schehr, P. Vivo, {\it  Phase transitions and edge scaling of number variance in Gaussian random matrices}, Phys. Rev. Lett. {\bf 112}, 254101 (2014).

\bibitem{Ricardo_PRE}
R. Marino, S. N. Majumdar, G. Schehr, P. Vivo, {\it Number statistics for $\beta$-ensembles of random matrices: applications to trapped fermions at zero temperature}, Phys. Rev. E {\bf 94}, 032115 (2016).





\bibitem{Eis13}
V. Eisler, {\it  Universality in the full counting statistics of trapped fermions}, Phys. Rev. Lett. {\bf 111}, 080402 (2013).	



\bibitem{CL95}
O. Costin, J.~L. Lebowitz, {\it  Gaussian fluctuation in random matrices}, Phys. Rev. Lett. {\bf 75},  69 (1995).


\bibitem{HC00}
T. L. Ho, C. Ciobanu, {\it  Rapidly rotating fermi gases}, Phys. Rev. Lett. {\bf 85}, 4648 (2000).

\bibitem{Ho01} T. L. Ho, {\it  Bose-Einstein condensates with large number of vortices}, Phys. Rev. Lett {\bf 87}, 060403 (2001).

\bibitem{ABD05}
A. Aftalion, X. Blanc, X., J. Dalibard, {\it  Vortex patterns in a fast rotating Bose-Einstein condensate}, Phys. Rev. A {\bf 71}, 023611 (2005).

\bibitem{TWC06}
G. Tonini, F. Werner, Y. Castin, {\it  Formation of a vortex lattice in a rotating BCS Fermi gas}, Eur. Phys. J. D {\bf 39}, 283 (2006).








\bibitem{SCEMC04}
V. Schweikhard, I. Coddington, P. Engels, V. P. Mogendorff, E. A. Cornell, {\it  Rapidly rotating Bose-Einstein condensates in and near the lowest Landau level}, Phys. Rev. Lett. {\bf 92}, 040404 (2004).

\bibitem{ZASSK05}
M. W. Zwierlein, J. R. Abo-Shaeer, A. Schirotzek, C. H. Schunck, W. Ketterle, {\it  Vortices and superfluidity in a strongly interacting fermi gas}, Nature  {\bf 435}, 1047 (2005).

\bibitem{Coo08}
For a review see N. R. Cooper, {\it  Rapidly rotating atomic gases}, Adv. Phys. {\bf 57}, 539 (2008).

\bibitem{Coo10}
N. R. Cooper, {\it Quantum Hall states of ultra cold atomic gases}, in {\it Many-Body physics with ultra cold gases}, Les Houches 2010, Eds. C. Salomon, G. Shlyapnikov, L. F. Cugliandolo, (2010). 

\bibitem{Ginibre}  
J. Ginibre, {\it  Statistical ensembles of complex, quaternion, and real matrices}, J. Math. Phys. {\bf 6}, 440 (1965).


\bibitem{Mehta}
M.~L. Mehta, {\it Random Matrices}, 2nd Edition, Academic Press (1991). 



\bibitem{Forrester_book}
P.~J. Forrester, {\it Log-gases and random matrices}, Princeton University Press, Princeton, NJ, (2010).






\bibitem{Landau}
L. D. Landau, E. M. Lifshitz, {\it Statistical Physics Part 1}, Volume 5, Butterworth Heinemann, Oxford (1981). 

\bibitem{Leggett}
A.~J. Leggett, {\it Quantum liquids: Bose condensation and Cooper pairing in condensed-matter systems}, Oxford University Press (2006).






\bibitem{GNV02}
A.~M. Garcia-Garcia, S.~M. Nishigaki, J.~J.~M. Verbaarschot, {\it Critical statistics for non-Hermitian matrices}, Phys.Rev. E {\bf 66}, 016132 (2002). 








\bibitem{Girko} 
V.~L. Girko, {\it  Circular law}, Theory Probab. Appl. {\bf 29}, 694 (1984)  







	%

  

  %

	



\bibitem{Supp_mat}
See supplementary material.

\bibitem{note_moy}

Note that the first cumulant is the mean number of eigenvalues $\moy{N_r}$.
It is obtained by integrating the mean density $\rho_N(r)$ on the disk ${\cal D}_r$.
In the deep and extended bulk, it converges to $r^2$. 
At the edge, we show that it converges to $\moy{N_r}=N+\sqrt{2N}{\cal K}_1^{\rm e}(\sqrt{2}(r-\sqrt{N}))$ with ${\cal K}_1^{\rm e}(s)=-\int_s^{\infty}dx \erfc(x)/2$.

\bibitem{EE_LLL}
I.~D. Rodr\'iguez, G. Sierra, {\it  Entanglement entropy of integer quantum Hall states}, Phys. Rev. B {\bf 80}(15), 153303 (2009).



\bibitem{foot_polylog}
Note that ${\rm Li}_{1-p}(-\erfc(x)/\erfc(-x))$ is a polynomial of order $p$ in the variable $\erf{x}$ \cite{Supp_mat}. 






\bibitem{gamma_exp}
NIST Digital Library of Mathematical Functions, \url{http://dlmf.nist.gov/8.11.E10}.


\bibitem{CMV12_2}
P. Calabrese, M. Mintchev, E. Vicari, {\it  Entanglement entropies in free-fermion gases for arbitrary dimension}, Europhys. Lett. {\bf 97}, 20009 (2012).



\bibitem{Area_law_1}
F.~G. Brand\~ao, M. Horodecki, {\it  An area law for entanglement from exponential decay of correlations}, Nat. Phys. {\bf 9}(11), 721 (2013).

\bibitem{Area_law_d_2}
M.~M. Wolf, F. Verstraete, M.~B. Hastings, J.~I. Cirac, {\it  Area laws in quantum systems: mutual information and correlations}, Phys. Rev. Lett. {\bf 100}(7), 070502 (2008).

\bibitem{Area_law_d}
J. Eisert, M. Cramer, M.~B. Plenio, {\it  Colloquium: Area laws for the entanglement entropy}, Rev. Mod. Phys. {\bf 82}(1), 277 (2010).


\bibitem{LGMS}
B. Lacroix-A-Chez-Toine, A. Grabsch, S.~N. Majumdar, G. Schehr, {\it Extremes of $2d$ Coulomb gas: universal intermediate deviation regime}, J.~Stat.~Mech. P013203, (2018).

\bibitem{foot_ddp}
Note that this result can also be obtained by using the determinantal structure of the Ginibre ensemble.  



%
%
%




\bibitem{wigner}
D. S. Dean, P. Le Doussal, S. N. Majumdar, G. Schehr, {\it Wigner function of noninteracting trapped fermions}, Phys. Rev. A {\bf 97}, 063614 (2018).

  %
  %
%
	%
%
  %
  %
  %
  %
  %
  %
%
%
  %
%
%
  %
  %
%
  %
  %
%
%
%
  %
  %
  %
  %
  %
  %
  %
%
%
%
%
  %
%
%
%
%
  %
%
  %
%
 %
 %
%
%
%
%
%
%
%
%
%
%
%
%
%
%
%




\end{thebibliography}

\begin{thebibliography}{}

\bibitem{gamma_exp}
NIST Digital Library of Mathematical Functions, \url{http://dlmf.nist.gov/8.11.E10}.

\bibitem{LGMS}
B. Lacroix-A-Chez-Toine, A. Grabsch, S.~N. Majumdar, G. Schehr, {\it Extremes of $2d$ Coulomb gas: universal intermediate deviation regime}, J.~Stat.~Mech. P013203, (2018).

\bibitem{erdelyi}
A. Erd\'elyi, W. Magnus, F. Oberhettinger, F. G. Tricomi, F.G. (1981), {\it Higher Transcendental Functions}, Vol. 1, Malabar, FL: R. E. Krieger Publishing (1981).

\bibitem{Forrester_book}
P.~J. Forrester, {\it Log-gases and random matrices}, Princeton University Press, Princeton, NJ, (2010).




\end{thebibliography}
\end{document}